# Near-infrared photoresponse in single walled carbon nanotube/polymer composite films

Biddut K. Sarker, M. Arif and Saiful I. Khondaker[*]
Nanoscience Technology Center and Department of Physics, University of Central Florida,
Orlando, Florida 32826, USA

**Abstract**

We present a near-infrared photoresponse study of single-walled carbon nanotube/poly(3-hexylthiophene)-block-polystyrene polymer (SWCNT/P3HT-b-PS) composite films for different loading ratios of SWCNT in the polymer matrix. Compared to the pure SWCNT film, the photoresponse [(light current – dark current)/dark current] is much larger in the SWCNT/polymer composite films. The photoresponse is up to 157% when SWCNTs are embedded in P3HT-b-PS while for a pure SWCNT film it is only 40%. We also show that the photocurrent strongly depends on the position of the laser spot with maximum photocurrent occurring at the metal–film interface. We explain the photoresponse due to exciton dissociations and charge carrier separation caused by a Schottky barrier at the metallic electrode - SWCNT interface.

1. **Introduction:**

Single walled carbon nanotubes (SWCNTs) have shown remarkable mechanical, optical and electrical properties making them particularly suitable for nanoelectronic and optical device applications [1-3]. However, the photoresponse in SWCNTs has generated considerable debate with the various studies leading to different interpretations about the origin of photoconductivity. In individual semiconducting SWCNT field effect transistor (FET) devices, photoresponse in the near-infrared (NIR) regime was explained using an exciton model [4,5]. In another study that involved individual SWCNT FET device, the photoresponse under UV illumination was explained by the desorption of molecular oxygen from the SWCNT surface which caused a reduction in hole carriers [6]. In contrast, for a large area SWCNT film, it was argued that the NIR photoresponse was caused by either the thermal effect [7, 8], or excitonic [9-12]. The ongoing debate about the origin of photoconductivity calls for further experimental and theoretical investigations not only in pure SWCNTs but also in materials where SWCNTs are used as filler such as SWCNT/polymer composite.

Incorporation of SWCNTs in the polymer matrix has lead to a new class of composite materials with multiple functionality and tailored properties [13-15]. For example, the electrical conductivity of SWCNT/polymer composite can be changed over 7/8 orders of magnitude by varying the concentration of SWCNTs in the polymer matrix. Optical absorption of SWCNTs are dominated by the singularities of their 1D band structure and absorb beyond the visible range into the NIR regime, while poly(3-hexylthiophene)-block-polystyrene polymer (P3HT-b-PS) do not show photo sensitivity beyond 700 nm. Therefore SWCNT/P3HT composite can absorb both in the visible and NIR regimes. In addition, their ease of processability in solution, mechanical flexibility, and low cost of device fabrication at macroscopic dimension make them attractive candidate for large area optoelectronic devices such as fast optical switches, photo detectors and solar cells. Despite their obvious advantages in optoelectronic applications, most of the previous studies on SWCNT/polymer composites focused on mechanical and/or electronic properties and the optoelectronic investigation of SWCNT/polymer composites did not receive much attention.

---

[*] Corresponding Author: Fax +1 407 882 2819. E-mail address: saiful@mail.ucf.edu (S. I. Khondaker)

In this paper, we present a photoresponse study of SWCNT/P3HT-b-PS polymer composite films with different SWCNTs loading ratios in the polymer matrix under NIR illumination. We show that the photoresponse [(light current – dark current)/dark current] or [$(I_{light}-I_{dark})/I_{dark}$] is up to 157% when SWCNTs are embedded in the polymer matrix while for pure SWCNT film it is only 40%. In addition, we found that the photocurrent in all of our composite films including the pure film is position dependent with maximum photocurrent occurring when illuminated at the metal-film interface and can be explained by excitonic model. This study not only provides further evidence about the origin of photoresponse in SWCNTs in favor of excitonic model but also offers opportunity to fabricate new classes of optoelectronic devices such as low cost infrared photo detectors and position sensitive detectors involving SWCNT/polymer composites.

## 2. Experimental details:

SWCNT/Polymer composites were prepared following our recently developed technique [14, 15]. In brief, SWCNTs were purchased from Carbon Nanotechnologies, Inc. In order to create a good dispersion of SWCNTs in the electrically and thermally insulating polystyrene (PS) matrix, first SWCNTs were dispersed into P3HT-b-PS in a 1:1 ratio. This is done by adding 5 mg of SWCNTs with 5 mg P3HT-b-PS into 5 ml chloroform in a vial followed by sonication for 1 hour in an ice water bath while the temperature was maintained at 18 -20 $^{0}$C. The SWCNTs were well dispersed and the solution was very uniform and stable. A 10 wt% PS solution was then made by adding 1.5 gm PS into 10 ml chloroform in a vial and placing it on a rotating stirrer for about 45 minutes. In order to make the composite solution, the appropriate amount of SWCNT/P3HT-b-PS mixture was added to the appropriate amount of PS solution. For example, 0.25% composite was made by mixing 0.25 ml solution of SWCNT/P3HT-b-PS with 1 ml solution of PS in a vial and placed on a shaker for about 7-8 minutes for fine mixing. To prepare the film, the resulting composite solution was spin coated at 300 rpm onto a cleaned glass substrate. Similarly, composite films were made with SWCNT weight percentage of 0.5%, 0.75%, 1%, 1.5%, 2%, 5%, 10%, in the polymer matrix. The average thickness of each film was about 60 μm. In addition, a pure P3HT-b-PS film and a pure SWCNT film were also prepared for control experiments. Finally, the conducting silver paint was used to make source and drain electrodes with separation of approximately 10 mm and width of 25 mm. The sample was then left at room temperature to dry for few hours. The degree of dispersion and orientation of SWCNTs in the polymer matrix were examined by high resolution Scanning Electron Microscopy (HRSEM, Zeiss ULTRA-55 FEG) at an accelerating voltage of 1 KV. Figure 1a shows the HRSEM image of a 1% composite film while figure 1b shows the HRSEM image of a pure SWCNT film. It can be seen from Figure 1a that SWCNTs are well dispersed in the polymer matrix and makes percolative pathways, which has been discussed in detail in our previous publications [14, 15].

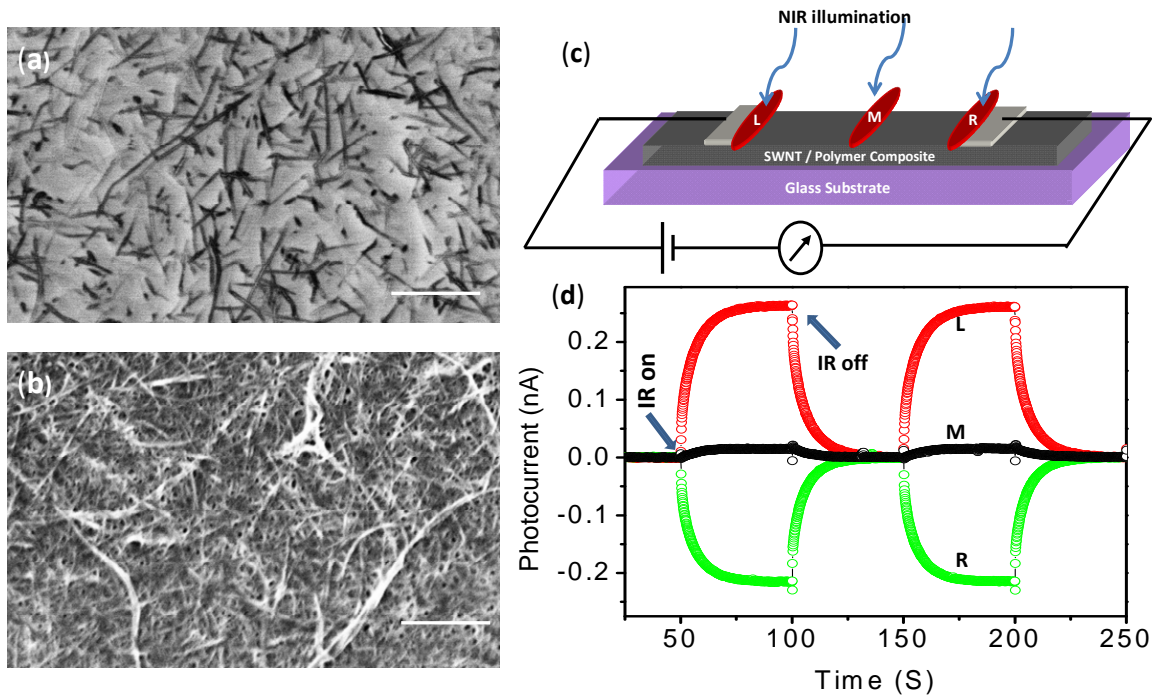

Figure 1: High Resolution Scanning electron Microscopy image of the (a) 1% wt SWCNT / P3HT-b-PS composite film (b) 100% SWCNT film (scale bar : 1 um). (c) Schematic diagram of the device and electric transport measurement set up. NIR laser wavelength was 808 nm and laser intensity was 4 mW/mm$^2$. L, M and R mark the positions of the laser with respect to the electrode. (d) Photocurrent as a function of time for 1% SWCNT/P3HT-b-PS composite film under NIR illumination at positions L, M and R. The IR laser is turned on t = 50 s and turned off and on at every 50 s interval. $V_{bias}$ = 1 mV.

Figure 1c shows a schematic diagram of a final device and the electrical transport measurement setup. The room temperature dc transport measurements of the composite films were carried out using a standard two-probe technique, both in the dark and under illumination by a laser spot, positioned at three different locations (Figure 1c). L corresponds to illumination on the left electrode/film interface, M is between the electrodes in the middle of the sample, and R is the right electrode/film interface. The near IR photo source consists of a semiconductor laser diode with peak wavelength of 808 nm (1.54 eV) driven by a Keithley 2400. The laser spot size was approximately 10 mm long and 1 mm wide. The photo intensity was monitored with a calibrated silicon photodiode (Thorlabs S121 B). The power intensity of the laser was ~ 4 mW/mm$^2$ when it was placed 20 mm above the sample for photoresponse measurements. Photocurrent was measured under small $V_{bias}$ (1mV) unless mentioned otherwise. Data was collected by means of LabView interfaced with the data acquisition card and current preamplifier capable of measuring sub pA signals.

### 3. Results and discussion:

Figure 1d shows a typical photoresponse curve for one of the composite films with 1% SWCNTs, where the photocurrent ($I_{photo}$) was plotted as a function of time ($t$) when the laser spot was positioned at L, M and R. The laser was turned on at t = 50 s and was switched off and on at 50 s intervals. Two cycles are shown to demonstrate the reproducibility of the on-off current. The photocurrent was calculated by subtracting the dark current ($I_{dark}$) from the current under laser light ($I_{light}$). Three features can be noticed from this figure, (a) the photocurrent is positional dependent, (b) there is a large enhancement of photocurrent at the metal-film interface, and (c) the response time is rather slow. When illuminated at position L there is an increase in photocurrent, while at position M there is a very small photocurrent generation, whereas position R shows a decrease in photocurrent. The dark current for this sample was 0.245 nA and the current under illumination at position L was 0.499 nA, giving an enhancement of 104%.

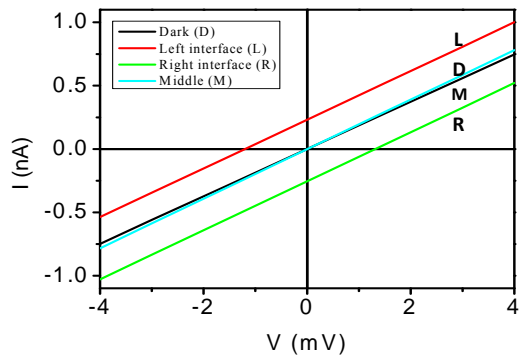

Figure 2: Current-voltage characteristics of a 1% SWCNT composite film in the dark and illuminated at position L, M and R on the sample.

In order to inspect the position dependent photocurrent further, we measured current–voltage (I –V) characteristics of the composite films in dark and under laser light illumination at positions L, M and R. Figure 2 shows a representative I-V curve for the 1% composite film. It can be seen that the I-V curves in the dark and at position M go directly though the origin, whereas when illuminated at position L and R, the I-V curve is slightly shifted above or below the origin, respectively. At zero applied bias, there is about +0.23 nA current at L and -0.26 nA at R. All the I-V curves show Ohmic behaviour from which we can calculate the resistances of the sample under different illumination condition. The resistances of the sample in the dark is 5.35 GΩ, while it is 5.11, 5.20 and 5.15 GΩ for NIR illumination at M, L and R respectively. So, the resistance slightly decreases under illumination with a maximum decrease of ~4.5%. A small change (less than 2%) in conductivity upon NIR illumination in SWCNT film was found in previous studies and was explained by a thermal effect. [7, 8] If thermal effect was responsible for photocurrent generation in our sample, it would have only generated a 4.5% change in photocurrent. Therefore, a photocurrent enhancement of 104% in our sample at the electrode-metal interface cannot be described using thermal effect. In addition, the resistance decreases under illumination at all positions which can cause an increase in photocurrent only. Whereas, in position R we observed a decrease in photocurrent providing further evidence that thermal effect is not responsible for photoresponse in our sample. Furthermore, a finite current at zero bias at position L and R suggests that a photovoltage is generated upon NIR illumination at the SWCNT-metal interface. Similar behaviour of the photocurrent has been observed in all our samples with different SWCNT loading ratios excluding the pure P3HT-b-PS film. For this sample, under 1 mV bias, signal to noise (S/N) ratio was extremely low, therefore the photocurrent was not very reproducible. The S/N ratios for all other composite films were more than 50 making the photocurrent easier to detection.

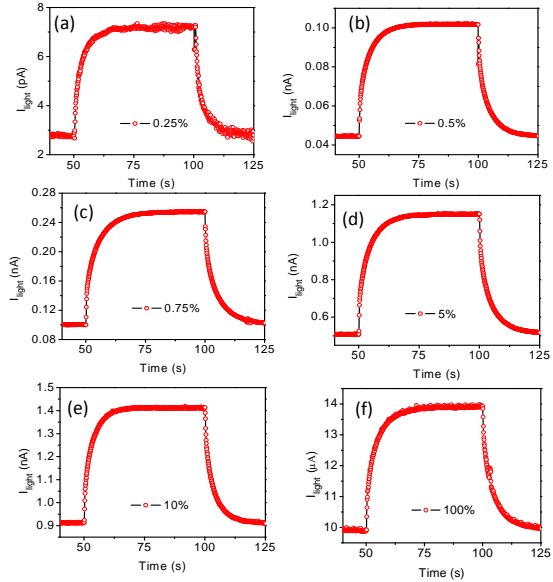

Figure 3: Photocurrent of (a) 0.25%, (b) 0.5%, (c) 0.75%, (d) 5%, (e) 10% SWCNT composite films and (f) pure SWCNT film. The NIR was turned on at t = 50 s and turned off at t = 100 s.

Table 1: Dark current, light current, photoresponse and external quantum efficiency (EQE) for SWCNT/P3HT-b-PS composite films under the NIR illumination at position L

| wt % of SWCNT | $I_{dark}$ (A) | $I_{light}$ (A) | $I_{photo}/I_{dark}$ (%) | EQE (%) |
|---|---|---|---|---|
| 0.25 | $2.80 \times 10^{-12}$ | $7.19 \times 10^{-12}$ | 156.78 | $1.7 \times 10^{-7}$ |
| 0.5 | $4.44 \times 10^{-11}$ | $1.01 \times 10^{-10}$ | 127.47 | $2.2 \times 10^{-6}$ |
| 0.75 | $1.00 \times 10^{-10}$ | $2.52 \times 10^{-10}$ | 152.00 | $5.8 \times 10^{-6}$ |
| 1 | $2.45 \times 10^{-10}$ | $4.99 \times 10^{-10}$ | 103.67 | $9.7 \times 10^{-6}$ |
| 1.5 | $3.38 \times 10^{-10}$ | $6.05 \times 10^{-10}$ | 78.99 | $1.0 \times 10^{-5}$ |
| 2 | $8.60 \times 10^{-10}$ | $1.82 \times 10^{-9}$ | 111.63 | $3.7 \times 10^{-5}$ |
| 5 | $5.09 \times 10^{-10}$ | $1.14 \times 10^{-10}$ | 123.97 | $2.4 \times 10^{-5}$ |
| 10 | $9.11 \times 10^{-10}$ | $1.41 \times 10^{-10}$ | 54.77 | $1.9 \times 10^{-5}$ |
| 100 | $9.95 \times 10^{-6}$ | $13.93 \times 10^{-6}$ | 40.00 | 0.15 |

Figure 3 shows $I_{light}$ for 0.25 %, 0.5%, 0.75%, 5%, 10% weight percentage of SWCNT in the polymer matrix and pure SWCNT film when illuminated at position L. The laser was switched on at t = 50 s and off at t = 100 s. Table 1 summarizes the experimental data from all of the samples, where dark current, light current, photoresponse, and external quantum efficiency (EQE) are shown for the different loading ratios of SWCNT composite films at position L. The EQE at λ = 808 nm at a fixed power of 4 mW/mm$^2$ was calculated using $EQE = (R_\lambda/\lambda) \cdot 1240 \ W \cdot nm/A$, where $R_\lambda$ is the responsivity defined by the photocurrent per watt of input power. As can be seen from Table 1 and Figure 3, the amount of SWCNTs in the polymer matrix has a strong influence on the photoresponse. The photoresponse is 157% for 0.25% SWCNT/P3HT-b-PS composite film while it is only 40% for pure SWCNT film. The stronger photoresponse in SWCNT/P3HT-b-PS films could be caused by the fact that the off current of the composite films can be reduced by several orders of magnitude by reducing the SWCNT content in the polymer matrix, which allows easy detection of the photocurrent when SWCNTs are in an electrically and thermally insulating polymer host. We note that in SWCNT-polycarbonate composite films studied by Pradhan et al. [8] a photoresponse enhancement in composite film was also observed. However, the maximum photoresponse in their study was only 5%. In our case, the maximum photoresponse is much higher (157%) which is 31 times larger than that of ref [8]. One possibility for the low enhancement in the study of Pradhan et al. [8], could be due to the fact that in their study, the illumination of NIR was around the middle part of the sample and that the authors did not check the effect of contact.

We now examine the slow time response of photocurrent when NIR source is switched on. Figure 4 is a plot of photocurrent as a function of time for the 1% composite film when the NIR is turned on at *t* =50 s, at position L (Figure 4a) and at position R (Figure 4b). When illuminated by the NIR source the current slowly increased (decreased for R) until it reached its steady state. We also measured the response of this film with a step function voltage (not shown here) to determine whether the slow time response had indeed come from the NIR illumination and not a delay due to a R-C like circuit existing in the entire setup. We found that unlike NIR source, the current increased almost instantaneously to a bias voltage switch. Therefore, the slow time response indeed comes from the NIR illumination and not a delay caused by a R-C like circuit existing in the entire setup. The dynamic response to the NIR source can be well described by $I = I_0[1 - \exp(-(t - t_0)/\tau)]$, where $\tau$ is time constant, $t_0$ is the time when NIR is switched on, and $I_0$ is the steady state photo current. In both figures, the open circles are the experimental data points and the solid lines are a fit to the above equation. From the fit in Figure 4 the time constant was calculated to be 5.7 and 4.9 seconds for position L and R respectively. Similar fits were also done for all the samples and time constants were calculated. The time constant measured for all our sample ranged from 3.18 to 5.7 s.

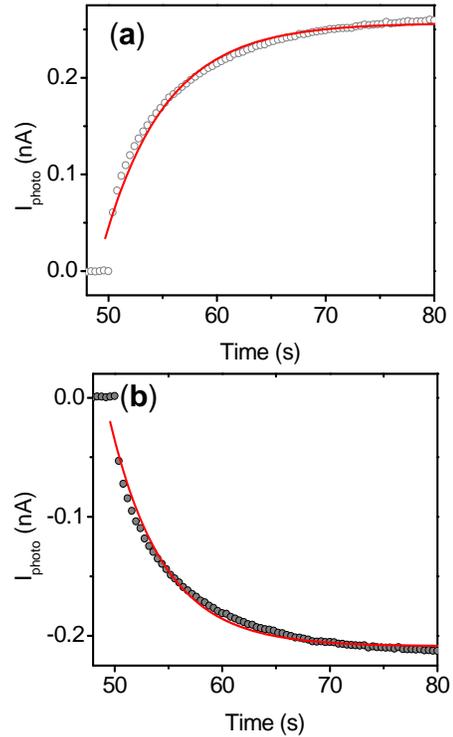

Figure 4. Time response of the photocurrent for 1% SWCNT/P3HT-b-PS composite film with NIR illuminated at position L (a) and R (b). The open circles are the experimental points and the solid line is an exponential fit of the data.

What causes photoresponse in our SWCNT/P3HT-b-PS composite samples? In previous studies, the photoresponse in SWCNT film has been explained by either thermal or excitonic mechanisms. In thermal mechanism, the temperature of the carbon nanotube film raises upon NIR illumination causing a decrease in the resistance of the nanotube film, hence a positive photoresponse [7]. We rule out thermal mechanism because both the positive and negative photoresponse at two different positions can not be explained by this model. In addition, the I-V curves at different positions show that even if thermal effect were present, it may account for a 5% change in current only, not the 157% seen in our sample. Our results presented here are consistent with the model of exciton generation upon NIR absorption and exciton dissociation at the metallic electrode -SWCNT interface due to a Schottky barrier [10]. In other words, when the laser light is illuminated at the interface, some energetic electrons overcome the asymmetric tunnel barrier at the interface and fall into the metal electrode leaving holes in the film. This causes an electron-hole separation at the interface and thereby creates a local electric field. Under the influence of this electric field, the carrier then diffuses to the other electrode through percolating SWCNT networks. Similar phenomenon occurs at the other electrode except that the right contact is a mirror image of the left contact and, therefore, the sign of photoresponse reverses. Whereas, when the laser is shined in the middle part of the sample electron hole pairs are created, however, the charge does not get separated so the overall photovoltage is almost zero. However, in our experiment, a very small photocurrent (5.7%) is seen at position M. The reason for a very small positive photocurrent at M can be explained as follows: the spot size of our near infrared (NIR) source is approximately 10 mm long and 1 mm wide and the positioning was done manually. Because of the finite width (10 mm long and 1 mm wide) of the NIR source and manual positioning, there is always a small error in positioning. Therefore although the laser was positioned in the middle, there might be small imbalance in positioning towards the left electrode which in turn can create charge carrier imbalance giving a very small photocurrent. Another reason could be that, even if the laser was positioned accurately at M, the thermal effect could also account for about 4.7% photocurrent. The interface between the SWCNTs and polymer may also help dissociate the exciton [16-18] and could be responsible for the ~ 5% change in conductivity in the middle part of the sample.

## 4. Conclusions:

In conclusion, we presented near infrared photoresponse study of SWCNT/P3HT-b-PS composite films for different loading ratios of SWCNT in the polymer matrix. We found that compared to the pure SWCNT film, the photoresponse was much larger when SWCNTs were embedded in the polymer matrix. The photoresponse was up to 157% for 0.5% composite film while for pure SWCNT film it was only 40%. In addition, we also found that photocurrent in all of our composite films were position dependent with maximum photoresponse occurring when illuminated at the metal-film interface. By comparing the magnitude of resistance changes from the current-voltage characteristic curves with that of the magnitude of the photoresponse, we conclude that the photoresponse was not caused by a thermal effect but by exciton dissociations at the metal-SWCNT film interface. This study show promises for SWCNT/polymer composite films for low cost infrared photo detectors and position sensitive detectors.

**Acknowledgements:**

We thank Dr. Jianhua Zou and Prof Lei Zhai for help with materials. This work is partially supported by US National Science Foundation under grant ECCS 0801924.